\documentclass[twocolumn,twocolappendix]{aastex631}

\begin{document}
\title{Detection of Enhanced Germanium in a New Cool Extreme Helium Star A\,980: Insights and Implications}

\correspondingauthor{Ajay Kumar Saini}
\email{ajay.saini@iiap.res.in}

\author[0009-0000-6371-907X]{Ajay Kumar Saini}
\affiliation{Indian Institute of Astrophysics, II Block, Koramangala, Bengaluru-560034, Karnataka, India}
\affiliation{Pondicherry University, R.V. Nagar, Kalapet, Pondicherry-605014, UT of Puducherry, India}

\author[0000-0001-5812-1516]{Gajendra Pandey}
\affiliation{Indian Institute of Astrophysics, II Block, Koramangala, Bengaluru-560034, Karnataka, India}



\begin{abstract}

A fine abundance analysis of a recently discovered hydrogen-deficient carbon (HdC) star, A\,980, is presented. Based on the observed high-resolution optical spectrum, we ascertain that A\,980 is a cool extreme helium (EHe) star and not an HdC. Singly-ionized germanium Ge\,{\sc ii} lines are identified in A\,980's optical spectrum. These are the first-ever detections of germanium lines in an EHe's observed spectrum, and provide the first measurements of germanium abundance in an EHe star. The overabundance of germanium in A\,980's atmosphere provides us with evidence for the synthesis of germanium in EHe stars. Among the known cool EHe stars, A\,980 exhibits a maximum enhancement of the $s$-process elements based on significant number of transitions. 
The measured elemental abundances reveal signs of H-burning, He-burning, and specifically the nucleosyntheses of the key elements: Ge, Sr, Y, Zr, and Ba. The nucleosyntheses of these key elements are discussed in light of asymptotic giant branch evolution and the expectation from the accretion of an He white dwarf by a C-O white dwarf or by a neutron star. 
\end{abstract}



\section{Introduction} \label{sec:intro}
There exists a distinct category of hydrogen-deficient stars that have enhanced carbon features and diminished hydrogen Balmer lines, in their observed spectra, than expected for their effective temperatures. This was first pointed out by \citet{1953ApJ...117...25B}. 

\citet{1953ApJ...117...25B} has listed four stars in this category, and these are not light or photometric variables\footnote{ Notably, there exists yet another class of hydrogen-deficient stars known as R Coronae Borealis (RCB) stars. RCB stars exhibit remarkable photometric variability by undergoing unpredictable light decline (up to about 9 mag in visual) in a matter of few weeks and recover to their maximum light in about few to several months \citep{1996PASP..108..225C}.}. \citet{1963MNRAS.126...61W} included one more star, HD\,148839, in this category of non-variables. Note that the observed spectrum of HD\,148839 shows the usual presence of enhanced C$_2$ bands and C\,{\sc i} lines, and the absence of CH band, except for not very weakened Balmer lines. 

Until the year 2022, there were just these five known in this category coined as hydrogen-deficient carbon (HdC) stars. Earlier, \citet{1967MNRAS.137..119W} conducted an abundance analysis of these five stars to investigate their surface composition. A recent survey has reported 27 new HdCs based on their low-resolution spectra - about a sixfold increase in their number than the earlier known \citep{2022A&A...667A..83T}. Detailed abundance analysis serves as a crucial observational constraint on theoretical models concerning the formation and evolution of these peculiar stars.

In this paper we conduct an abundance analysis of a new warm HdC, A\,980 (2MASS 18113561+0154326), using a high-resolution spectrum.
\section{Observations, Data Reduction, and Line Identification} \label{sec:observations}

We have obtained a high-resolution optical spectrum of the newly identified warm HdC star A\,980 using the Hanle Echelle Spectrograph (HESP) mounted on 2-m Himalayan Chandra Telescope (HCT) at the Indian Astronomical Observatory (IAO) in Hanle, Ladakh, India. These observations were made on 21 June 2023. The spectrum was recorded onto a 4k $\times$ 4k CCD that covered the wavelength range from 3856 \AA\ to 9656 \AA\ with a resolving power ($R = \lambda/\Delta\lambda$) of about 30\,000. A total of three frames, each of 45 minutes exposure, were observed and co-added to improve the signal-to-noise ratio (SNR) (ranging from 25 to 180 per pixel across the covered spectral range from blue to red). 
A Th-Ar hollow cathode lamp was observed for wavelength calibration. To normalize the pixel-to-pixel variation in the sensitivity of the CCD, several exposures known as flat frames with differing spectrograph focus (in focus and out of focus) were acquired using a featureless quartz-halogen lamp. All the flat frames were combined to create a master flat with a very high signal for flat correction. A telluric standard, that is a rapidly rotating B-type bright star, was also observed on the same night to get rid of Earth's atmospheric absorption lines. The observed spectrum of the star was then reduced utilizing the standard tasks under the Image Reduction and Analysis Facility (IRAF) software package \citep{1986SPIE..627..733T,1993ASPC...52..173T}. In IRAF, the task \texttt{telluric} is used to remove Earth's atmospheric absorption lines by dividing the observed telluric standard spectrum from the observed spectrum of our program star.

\begin{figure*}[ht!]
\includegraphics[width=\hsize]{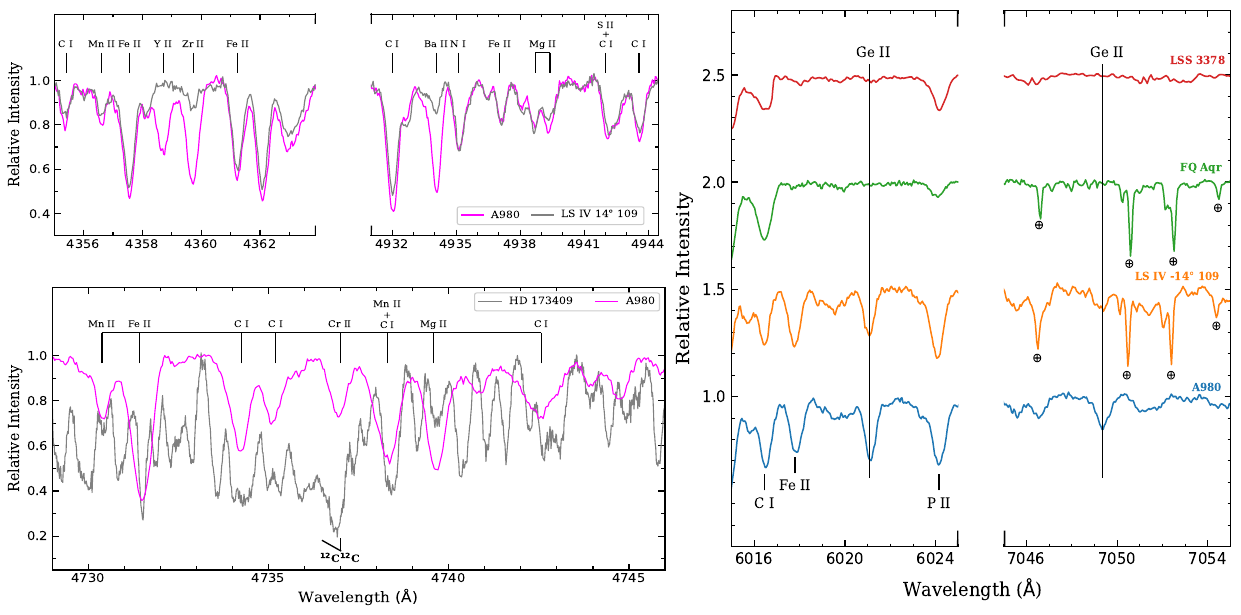}
\caption{Observed spectra of LS IV -14\textdegree\,109, a known cool EHe, and A\,980 are shown for comparison (see top left panel). Observed spectra of HD\,173409, a known HdC, and A\,980 in (1, 0) C$_2$ band region for comparison (bottom left panel). In the right panel, identification of Ge II $\lambda$6021.04 and 7049.37 \AA\ lines in A\,980 and other cool EHes LS IV -14\textdegree\ 109, FQ\,Aqr, and LSS\,3378 are highlighted by vertical lines. The principal lines are marked. The telluric absorption lines are shown by encircled crosses. The key is provided on each panel.
\label{fig:identification}}
\end{figure*}

For the spectral line identifications, we have used the NIST Atomic Spectra Database\footnote{see \url{https://physics.nist.gov/asd}}, A multiplet table of astrophysical interest - revised edition \citep{moore1972}, Tables of spectra of H, C, N, and O \citep{moore1993}, Kurucz’s database\footnote{see \url{http://kurucz.harvard.edu}} and VALD database\footnote{see \url{https://vald.astro.uu.se}}. A large spectral coverage (3856-9656 \AA) of the star's observed spectrum has enabled us to identify several important elements with their significant number of lines, of about 1900, and their ionization stages. The observed absorption line spectrum is well represented by neutral helium lines, neutral and singly ionized carbon lines, and also by plenty of singly ionized metal lines of the iron group\footnote{The radial velocity (R.V.) of A\,980, determined from the identified Fe\,{\sc ii} lines, is measured to be 109 $\pm$ 2 km s$^{-1}$.}. Singly ionized helium is not detected in the spectrum. Lines of all elements expected and observed in early A-type and late B-type normal stars are found. The observed spectrum exhibits remarkably diminished hydrogen Balmer lines and the absence of C$_2$ molecular features. We note that the above described characteristics fit well with the observed absorption lines of an extreme helium (EHe) star especially, the cooler EHes \citep{2001MNRAS.324..937P}. A comparison with the available high-resolution spectrum of the cool EHe, LS IV -14\textdegree\ 109, reveals that the observed spectrum of the program star A\,980 is a near replica of the former except for the significantly enhanced absorptions of neutron-capture elements in A\,980's spectrum (see Figure \ref{fig:identification}, top left panel, for example). The presence of He\,{\sc i} line in the observed spectrum of A\,980 is also shown along with that of LS IV -14\textdegree\ 109 in Figure \ref{fig:parameters}d. The observed spectrum of an HdC, HD\,173409 (available from \citet{2012ApJ...747..102H}) when compared with that of A\,980 shows the absence of the C$_2$ molecular features in A\,980 (see Figure \ref{fig:identification}, bottom left panel, for example). These comparisons clearly demonstrate that A\,980 is not an HdC star. However, \citet{2022A&A...667A..83T} have classified A\,980 as a warm HdC star, and have renamed the category of HdCs as dustless HdC (dlHdC) stars.

Among the several absorptions of the neutron capture elements in A\,980's spectrum, the notable were the identifications of singly-ionized germanium Ge\,{\sc ii} lines.

\subsection{Ge\,{\sc ii} lines} \label{sec:gelines}

Only two multiplets are listed in the Revised Multiplet Table (RMT; \citet{moore1972}) for the identification of Ge\,{\sc ii} lines. Out of the two Ge\,{\sc ii} lines of RMT\,1, the line at 6021.09\AA\ is clearly present in A\,980's spectrum while the stronger line at 5893.42\AA\ is severely blended as engulfed by a strong ISM-NaD$_1$ component. No Ge\,{\sc ii} lines of RMT\,2 are detected in the observed spectrum. However, the NIST Atomic Spectra Database provides more multiplets that include the above-discussed two multiplets listed in the RMT for further line identification. A thorough search was made for all Ge\,{\sc ii} lines of these multiplets. Two more unblended Ge\,{\sc ii} lines, $\lambda$7049.37 and $\lambda$7145.39, from multiplet $^2D-2P^0$ of transition array $4s4p^2 - 4s^25p$ were identified; the third line $\lambda$6966.32 of this multiplet is severely contaminated by a Fe\,{\sc ii} line. The Ge\,{\sc ii} line $\lambda$6021.09 that was identified using the RMT is from multiplet $^2S-2P^0$ of transition array $4s^25s - 4s^25p$. We note that Ge\,{\sc ii} lines from the multiplets of the excited levels with lower excitation potentials of 9.7 eV or higher have negligible contribution to the observed spectrum of A\,980. We also searched for Ge\,{\sc ii} lines in the available spectra of three cooler EHes: LS IV -14\textdegree\ 109, FQ\,Aqr, and LSS\,3378 (see \citet{2001MNRAS.324..937P};   \citet{2006MNRAS.369.1677P}). All the Ge\,{\sc ii} lines that were detected in A\,980's spectrum are also present in the available spectrum of LS IV -14\textdegree\ 109. Along with these detected lines, the spectrum of LS IV -14\textdegree\ 109 revealed an additional unblended line $\lambda$5893.39, the strongest line from the multiplet RMT\,1, $^2S-2P^0$, which was expected to be present in A\,980's spectrum but was severely contaminated as discussed earlier. Our search for all the
above identified Ge\,{\sc ii} lines, including the strongest line at 5893.39\AA, in the available spectra of the other two cool EHes: FQ\,Aqr and LSS\,3378 was unsuccessful (for example, see right panel of Figure \ref{fig:identification}).
\section{Abundance Analysis - Procedure}
The procedure for abundance analysis involves predicting the observed spectrum using a grid of model atmospheres combined with a radiative transfer code in local thermodynamic equilibrium (LTE). The predictions of an observed spectrum were made in the sense of computing the line equivalent width or synthesizing the line spectrum. 

\subsection{Hydrogen-deficient model atmospheres}
For the analysis of A\,980, a cool EHe star, the same procedure is followed as described in \citet{2001MNRAS.324..937P}. The abundance analysis uses two grids of line-blanketed hydrogen-deficient model atmospheres; one with $T_{\text{eff}}\leq 9500$ K and the other with $T_{\text{eff}}\geq 10\ 000$ K. The grid of $T_{\text{eff}}\leq 9500$ K, the Uppsala line-blanketed models,
is described by \citet{1997A&A...318..521A}, and that of $T_{\text{eff}}\geq 10\ 000$ K was computed by the code STERNE \citep{1992A&A...260..133J}. The line formation calculations, which involve computing the line equivalent width as well as the spectrum synthesis using the Uppsala models, were carried out with the Uppsala LTE line-formation code EQWIDTH. Similarly,
spectrum synthesis code SYNSPEC \citep{2017arXiv170601859H} was adopted for the line formation calculations using STERNE as well as Uppsala models.

\subsection{Consistency between line-formation codes, and between model grids}
For an adopted model atmosphere, the abundances derived using SYNSPEC and EQWIDTH for weak lines are in good agreement within 0.1 dex for a majority of the species. We notice that the abundances derived using the former are always lower than those derived using the latter. We attribute these small differences to the adopted continuous opacity data being from two different sources.  

The two model atmosphere grids, as discussed above, do not overlap in effective temperature. These two grids are compared by deriving a model for 9500 K using extrapolation of the high-temperature grid whose coolest models are at 10\,000 K. The extrapolated model and an Uppsala model for 9500 K gave identical abundances to within 0.05 dex \citep{2001MNRAS.324..937P}.

\subsection{Atomic data}
The $gf$-values and excitation potentials for most of the lines used in our LTE analysis were available in the NIST database.  The data for the remaining lines are from Kurucz database and other sources (see appendix: Table \ref{tab:linelist}). However, for several Zr\,{\sc ii} lines, we have used new improved experimental $gf$-values from the compilations of \cite{2006A&A...456.1181L} and \cite{2006MNRAS.367..754M}. The Stark broadening and radiation
broadening coefficients are mostly from the Kurucz database. The line broadening coefficients for computing the He\,{\sc i} profiles are from \cite{1984JQSRT..31..301D}. The line broadening treatments for synthesizing hydrogen Balmer lines are from \cite{1994A&A...282..151H} as used by SYNSPEC. See appendix: Table \ref{tab:linelist} for the detailed line list used in our analysis.

\subsection{Atmospheric parameters and elemental abundances}
A model atmosphere is characterized by its effective temperature, surface gravity, and chemical composition. The input elemental abundances of the adopted model atmosphere need to be consistent with those derived from the observed spectrum. A model atmosphere of an EHe star is mainly governed by the input carbon and helium abundances. Specifically, both carbon and helium are the major contributors to the continuum opacity in cool EHes \citep{2001MNRAS.324..937P}. Hence, the input C/He ratio of the adopted model atmosphere needs to be in accord with that determined from the star's observed spectrum. The input composition of the rest of the elements is scaled to solar with H/He fixed at 10$^{-4}$ by number.

A fine abundance analysis starts with deriving the star's effective temperature (T$_{\text{eff}}$), surface gravity (log\,g), and the microturbulent velocity ($\xi$). Subsequently, the photospheric elemental abundances are estimated for the adopted model atmosphere. These parameters are determined from the observed line spectrum. 

The microturbulent velocity $\xi$ (in km s$^{-1}$) is first determined by imposing that the lines of a particular species, neutral or ionized, with similar lower excitation potential (LEP) and having a range in their measured equivalent widths, return the same abundance. 

Second, the condition of the ionization balance needs to be satisfied i.e., for an element represented in the observed spectrum by neutral as well as the ionized states must return the same abundance from lines of different stages of ionization. This designates a locus in the (T$_{\text{eff}}$, log\,g) plane. Such elements with their pairs of ions provide loci of very similar slope in the (T$_{\text{eff}}$, log\,g) plane. 

The third condition that needs to be satisfied is the excitation balance which serves as a thermometer measuring T$_{\text{eff}}$. This requires that the lines of a particular species but of differing LEPs should return the same elemental abundance. The model grid is searched for the model that meets this condition. The optimum T$_{\text{eff}}$ is found to be independent of the adopted log\,g and C/He for the model atmosphere. Hence, this indicator provides a locus in the (T$_{\text{eff}}$, log\,g) plane to lift the degeneracy presented by the ionization balance. A second indicator may be available: for stars hotter than about 10\,000 K, the He\,{\sc i} profiles are less sensitive to T$_{\text{eff}}$ than to log\,g on account of pressure broadening due to the quadratic Stark
effect.  

In reproducing the observed spectrum by spectrum synthesis, we include broadening due to the instrumental line profiles, the microturbulent velocity $\xi$ and assign all additional broadening, if any, to rotational broadening. Observed unblended line profiles, preferably weak, are used to obtain the additional broadening. Synthetic line profile, including the broadening due to instrumental profile, for the adopted model atmosphere (T$_{\text{eff}}$, log\,g, $\xi$) and the abundance (from equivalent width analysis), is found to be sharper than the observed line profiles. This extra broadening in the observed profile is attributed to rotational broadening.
\section{A\,980 $-$ Abundance analysis results}
\begin{figure*}[ht!]
\includegraphics[width=\hsize]{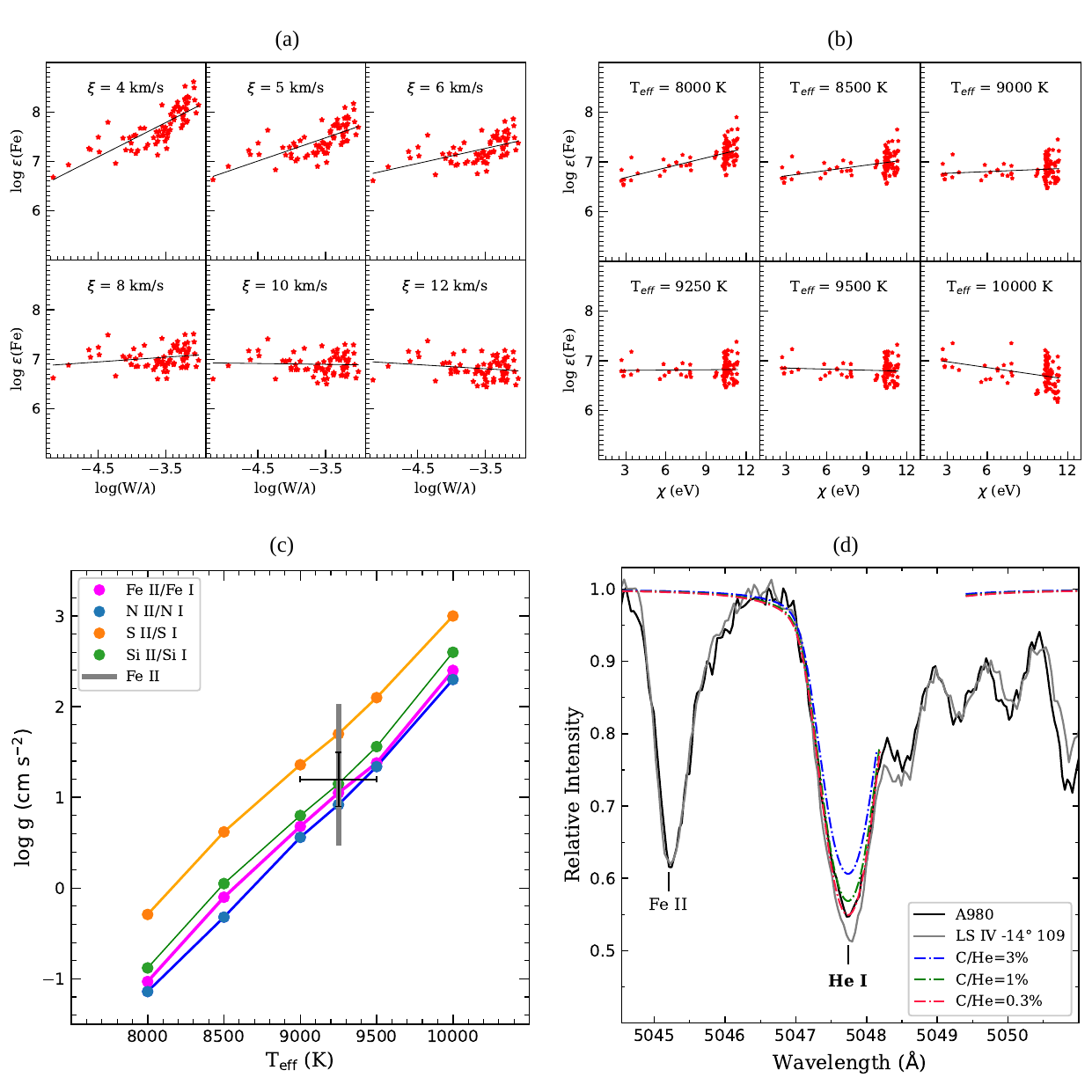}
\caption{(a) Abundances from Fe\,{\sc ii} lines for A\,980 versus their reduced equivalent widths (log $W_\lambda/\lambda$). A value of $\xi$ = 11.0 km/s is obtained from this figure. (b) Abundances from Fe\,{\sc ii} lines for A\,980 versus their lower excitation potential ($\chi$), showing excitation equilibrium. A value of T$_{\text{eff}}$ = 9250 $\pm$ 250 K is obtained from this figure. (c) The (T$_{\text{eff}}$, log\,g) plane of A\,980 shows the final values of T$_{\text{eff}}$ and log\,g with error bars. Loci satisfying ionization
equilibria are plotted $-$ see keys on the figure. The excitation equilibrium of Fe\,{\sc ii} lines is shown by thick \& solid gray line. (d) Synthesized and observed He\,{\sc i} $\lambda$5047.74 \AA\ profiles. The He\,{\sc i} line profiles are synthesized using the model T$_{\text{eff}}$ = 9250 K, log\,g = 1.25 cgs and $\xi$ = 11.5 km s$^{-1}$ for three different values of C/He values. The observed He\,{\sc i} $\lambda$5047.74 \AA\ profile of LS IV -14\textdegree\,109 is also shown for comparison. The key is provided on each panel.
\label{fig:parameters}}
\end{figure*}
A test model atmosphere is adopted with (T$_{\text{eff}}$ = 9500 K, log\,g $=$ 1.0, C/He $=$ 1\%) to first determine the microturbulent velocity $\xi$. These stellar parameters were chosen due to the near identical observed spectra of A\,980 and LS IV -14\textdegree\ 109 (see Section \ref{sec:observations}). The stellar parameters derived for LS IV -14\textdegree\ 109 by \citet{2001MNRAS.324..937P} are the same as adopted above. $\xi$ is determined from lines of Fe\,{\sc ii}, Ti\,{\sc ii}, Cr\,{\sc ii}, and C\,{\sc i}. $\xi$ is found to be 11.0 km s$^{-1}$ from Fe\,{\sc ii} and Ti\,{\sc ii} lines while that from Cr\,{\sc ii} and C\,{\sc i} lines is 12.0 km s$^{-1}$ and 11.5 km s$^{-1}$, respectively. We finally adopt a mean $\xi$ of 11.5 $\pm$ 1.0 km s$^{-1}$ for determining the elemental abundances. For example, see Figure \ref{fig:parameters}a that illustrates the procedure of estimating $\xi$. 

T$_{\text{eff}}$ was obtained from Fe\,{\sc ii} lines that exhibit a range in their lower excitation potentials (LEP): 2 to 11 eV. For $\xi =$ 11.5 km s$^{-1}$, we found a model (T$_{\text{eff}}$, log\,g, $\xi$) that provided the same abundance independent of the line's LEP. The T$_{\text{eff}}$ deduced from the Fe\,{\sc ii} lines has negligible dependence on the adopted model's log\,g and C/He ratio.
Figure \ref{fig:parameters}b illustrates the procedure for obtaining T$_{\text{eff}}$. The loci obtained from Fe\,{\sc ii}/Fe\,{\sc i}, N\,{\sc ii}/N\,{\sc i}, Si\,{\sc ii}/Si\,{\sc i}, and S\,{\sc ii}/S\,{\sc i} ionization balance are shown in Figure \ref{fig:parameters}c. These combined with the T$_{\text{eff}}$ deduced from the excitation balance of Fe\,{\sc ii} lines, that has negligible dependence on surface gravity, eventually provide the optimum stellar parameters: T$_{\text{eff}}$ = 9250 $\pm$ 250 K, log\,g = 1.2 $\pm$ 0.3 cm s$^{-2}$ (cgs), and $\xi$ = 11.5 $\pm$ 1.0 km s$^{-1}$. 

For the derived stellar parameters, the C/He ratio may directly be determined from the measured equivalent widths of C\,{\sc i} and C\,{\sc ii} lines. The observed He\,{\sc i} line profiles are presumably yet another indicator of the C/He ratio. Figure 5 of \citet{2001MNRAS.324..937P} illustrates the predicted equivalent widths of C\,{\sc i}, C\,{\sc ii}, and He\,{\sc i} lines for a model's C/He ratio. At the derived T$_{\text{eff}}$ and log\,g of A\,980, the predicted C\,{\sc i} line strengths are insensitive to the model's C/He $\ge 1.0\%$; C/He of 1\% corresponds to carbon abundance, log $\epsilon$(C) $= 9.5$. However, the C\,{\sc ii} line strengths are nearly insensitive to the model's C/He $\ge 0.3\%$. The observed C\,{\sc ii} lines are very strong and the measured equivalent widths are very large. Hence, we conclude that C\,{\sc ii} lines are not fit for estimating the C/He ratio and instead use C\,{\sc i} lines for determining the same. 

The C\,{\sc i} lines return carbon abundances of log $\epsilon$(C) = 8.9 $\pm$ 0.2, 9.0 $\pm$ 0.2, and 9.3 $\pm$ 0.2 for the optimum model's input ratio of C/He $=$ 0.3\% (log $\epsilon$(C) = 9.0), 1.0\% (log $\epsilon$(C) = 9.5), and 3.0\% (log $\epsilon$(C) = 10.0), respectively. We find that the input model's carbon abundance is consistent with that derived for the optimum model with a C/He of 0.3\%. He\,{\sc i} profiles at 5048, 5876, and 6678 \AA\ may provide an estimate of the C/He ratio. The best fitting observed profile of He\,{\sc i} is shown in Figure \ref{fig:parameters}d. A ratio C/He of about 0.3\% produces an acceptable fit to the He\,{\sc i} profile. Within the uncertainties, mainly line-to-line scatter, a ratio C/He $=$ 0.3\% is determined from the C\,{\sc i} lines as well as the He\,{\sc i} profiles independently.

\begin{figure*}[t]
\includegraphics[width=\hsize]{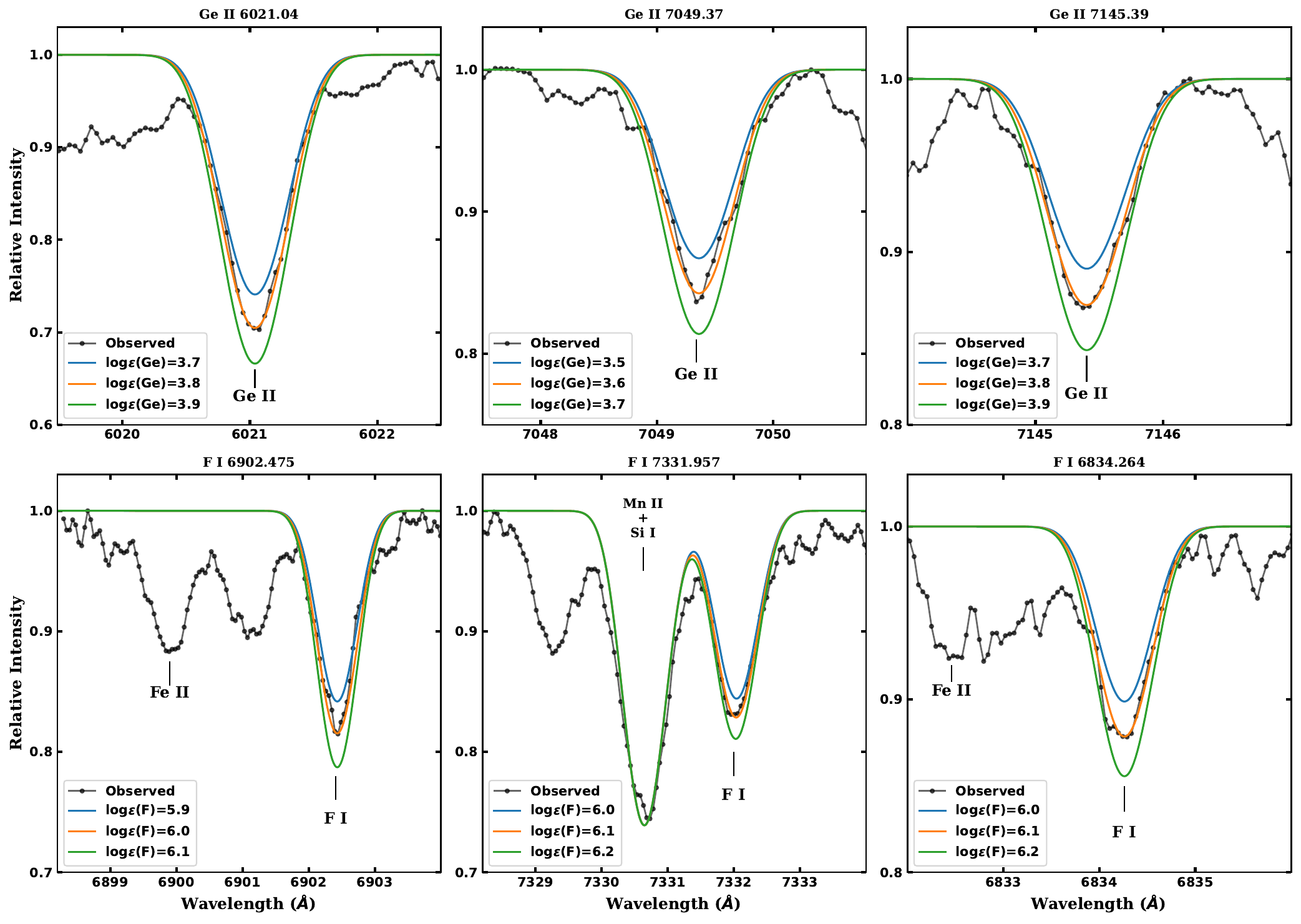}
\caption{For A\,980, top and bottom panels show the spectrum syntheses of Ge\,{\sc ii} and F\,{\sc i} lines, respectively. See key on the Figures.
\label{fig:gefsyntheis}}
\end{figure*}

The final abundances, as given in Table \ref{tab:abundancetable}, are derived for the optimum model of C/He $=$ 0.3\%. The derived abundances are normalized based on the convention that $\log\epsilon$(X) = $\log(X/H)$ + 12.0 to a scale in which $\log\sum\mu_I\epsilon(I)$ = 12.15, where 12.15 is determined from solar abundances with He/H $\simeq$ 0.1. Here, $\mu_{\rm X}$ is the atomic weight of element X. Since all elements but He have a very low abundance, the helium abundance $\log\epsilon$(He) is 11.54 on this scale. The hydrogen abundance is from the syntheses of H$\alpha$, H$\beta$, and H$\gamma$ profiles, for example, see appendix: Figure \ref{fig:Hsyntheis}. Neutral lines of the key element fluorine are also present in the spectrum of A\,980 like other cool EHes \citep{2006ApJ...648L.143P}. Fluorine abundance is determined by spectrum syntheses of neutral fluorine (F\,{\sc i}) lines (see Figure \ref{fig:gefsyntheis}, bottom panels). The abundance of germanium, a very significant element for this study, is from spectrum syntheses of Ge\,{\sc ii} lines at 6021.04, 7049.37, and 7145.39\AA\ (see Figure \ref{fig:gefsyntheis}, top panels). The deduced projected rotational velocity (\textit{v}\,sin\textit{i}) of about 15 km s$^{-1}$ is adopted for spectrum syntheses.
\section{Surface composition $-$ Discussion}
The derived surface composition of A\,980 is summarized in Table \ref{tab:abundancetable} along with the comparison star LS IV -14\textdegree\,109 \citep{2001MNRAS.324..937P}; also given are the solar abundances from \cite{2009ARA}. The derived abundances of the cool EHe A\,980 are then compared with the measured abundances of other EHes from the literature (see \citet{2021ApJ...921...52P} and references therein).

The surface composition of A\,980 clearly reveals that its atmosphere is hydrogen poor and rich in helium and carbon as observed for most of the EHes. These are clues that the atmosphere of A\,980 is contaminated by the products of H- and He- burning reactions.
\begin{deluxetable}{ccccccc}
\tablecolumns{7}
\tablewidth{\hsize}
\tablecaption{Derived elemental abundances for A\,980 and LS IV -14\textdegree\ 109. \label{tab:abundancetable}}
\tablehead{
\colhead{Element} & \colhead{Solar\tablenotemark{a}} & \multicolumn{2}{c}{A\,980\tablenotemark{b}} & \multicolumn{2}{c}{LS IV -14\textdegree\ 109\tablenotemark{c}} \\
\colhead{} & \colhead{log $\epsilon$(X)} & \colhead{log $\epsilon$(X)} & \colhead{[X/Fe]} & \colhead{log $\epsilon$(X)} & \colhead{[X/Fe]}
}
\startdata
H & 12 & 5.81 & -5.39 & 6.20 & -5.30 \\
He & 10.93 & 11.54 & 1.41 & 11.54 & 1.11 \\
C & 8.43 & 8.87 & 1.24 & 9.45 & 1.52 \\
N & 7.83 & 8.40 & 1.37 & 8.60 & 1.27 \\
O & 8.69 & 7.89 & 0.00 & 8.50 & 0.31 \\
F & 4.56 & 6.10 & 2.34 & 6.52 & 2.46 \\
Ne & 7.93 & 8.65 & 1.52 & 9.40 & 1.97 \\
Na & 6.24 & 6.46 & 1.02 & 6.80 & 1.06 \\
Mg & 7.60 & 7.00 & 0.20 & 7.20 & 0.10 \\
Al & 6.45 & 6.14 & 0.49 & 6.90 & 0.95 \\
Si & 7.51 & 7.30 & 0.59 & 7.68 & 0.67 \\
P & 5.41 & 5.36 & 0.75 & 5.30 & 0.39 \\
S & 7.12 & 6.89 & 0.57 & 7.55 & 0.93 \\
Ar & 6.40 & 5.99 & 0.39 & ... & ... \\
Ca & 6.34 & 5.64 & 0.10 & 5.55 & -0.29 \\
Sc & 3.15 & 2.98 & 0.63 & 3.30 & 0.65 \\
Ti & 4.95 & 4.27 & 0.12 & 4.30 & -0.15 \\
V & 3.93 & 3.40 & 0.27 & ... & ... \\
Cr & 5.64 & 4.75 & -0.09 & 5.10 & -0.04 \\
Mn & 5.43 & 4.85 & 0.22 & 5.30 & 0.37 \\
Fe & 7.50 & 6.70 & 0.00 & 7.00 & 0.00 \\
Co & 4.99 & 4.72 & 0.53 & ... & ... \\
Ni & 6.22 & 5.50 & 0.08 & 5.92\tablenotemark{d} & 0.20 \\
Ge & 3.65 & 3.72 & 0.87 & 3.48\tablenotemark{b} & 0.33 \\
Sr & 2.87 & 3.40 & 1.33 & 2.60 & 0.23 \\
Y & 2.21 & 2.54 & 1.13 & 1.90 & 0.19 \\
Zr & 2.58 & 2.93 & 1.15 & 1.90 & -0.18 \\
Ba & 2.18 & 2.23 & 0.85 & 1.70 & 0.02 \\
\enddata
\tablenotetext{a}{Solar abundances are adopted from \cite{2009ARA}}
\tablenotetext{b}{This work}
\tablenotetext{c}{Adopted from \cite{2001MNRAS.324..937P} and \cite{2006ApJ...648L.143P}}
\tablenotetext{d}{From the unblended cleaner lines and their adopted $gf$-values used in this work for A\,980}
\tablecomments{[X/Fe] = (X/Fe)$_{\star}$ $-$ (X/Fe)$_{\odot}$}
\end{deluxetable}
For the evolved low- and intermediate-mass stars, abundances of elements that are unlikely to be affected by H- and He-burning and attendant nuclear reactions are good indicators of their initial metallicity. These are V, Cr, Mn, Fe, Co, and Ni $-$ the iron-peak elements, and Mg, Si, S, Ca, Sc, and Ti $-$ the $\alpha$-elements. We note that A\,980's abundance ratios, Mg/Fe, Si/Fe, S/Fe, Ca/Fe, Sc/Fe, V/Fe, Cr/Fe, Mn/Fe, Co/Fe and Ni/Fe, are on average as expected for that in metal-poor normal and unevolved stars (see Table \ref{tab:abundancetable}; \citet{2000A&A...359..191G,2004A&A...415..559R}). We adopt iron abundance as the indicator of initial metallicity in A\,980.

The abundances of elements that are affected in the course of A\,980's evolution are H, He, C, N, O, F, Ne, Ge, Sr, Yr, Zr, and Ba when compared to a normal star, for example, the Sun $-$ see Table \ref{tab:abundancetable}.

We discuss by comparing the derived surface composition of A\,980 with other EHes. The surface composition of other EHes including the hot RCB star DY\,Cen are from
\citet{1992A&A...260..133J,1998A&A...329.1019D,jeffery1998,1998A&A...340..476J,2001MNRAS.324..937P,2004ApJ...602L.113P,2006ApJ...638..454P,2006ApJ...648L.143P,2006MNRAS.369.1677P,2011ApJ...727..122P,2014ApJ...793...76P,2017ApJ...847..127P,2017MNRAS.470.3557J,2020ApJ...891...40B,2024MNRAS.530.1666J}.

\textit{Hydrogen} -- The hydrogen abundance log $\epsilon$(H) is found to be 5.8 $\pm$ 0.2 which is under-abundant by a factor of $\sim10^6$ relative to a normal star. A\,980's hydrogen abundance fits well within the range of 5$-$8 that is derived for other EHes (see \citet{2021ApJ...921...52P} and references therein). The exceptions are the hot RCB, DY\,Cen, and the EHes with a very low carbon abundance (see the following section).

\textit{Carbon} -- The carbon abundance log $\epsilon$(C) is found to be 8.9 $\pm$ 0.2 that is a C/He ratio of 0.0023. A\,980 falls at the lower end of the range in the C/He ratios for other EHes. Here we exclude the four known EHes with a very low carbon abundance. However, recent studies suggest that HD\,144941, one of these four, is the most extreme helium-strong star with extremely strong magnetic field (see \citet{2021RNAAS...5..254P,2021MNRAS.507.1283S}).

\textit{Nitrogen} -- Nitrogen abundance is enriched by a factor of 25 (1.4 dex) relative to the expected initial abundance for A\,980's metallicity that is the Fe abundance. The enhanced nitrogen abundance shows the complete conversion of initial C, N, and O to N. This provides evidence for the operation of severe CNO cycling in an H-rich region. 

\textit{Oxygen} -- Oxygen abundance is close to the initial value expected for A\,980's Fe abundance. The O/N ratio is found to be 0.94 for A\,980 and is in concert with the other EHes having an O abundance close to their initial value (see \citet{2006ApJ...638..454P}).

\textit{Fluorine} -- Fluorine abundance is significantly enhanced that is about a factor of 200 when compared with the F expected for A\,980's metallicity. This enhancement is in concert with other cool and hot EHes including the RCB stars (see \citet{2020ApJ...891...40B}).

\textit{Neon} -- Neon abundances are from Ne\,{\sc i} lines, and are subject to non-LTE effects. Application of the non-LTE effects most likely brings the Ne abundances of the cool EHes, here A\,980, (down by about 0.8 dex) in line with the hot EHes \citep{2011ApJ...727..122P}.

\textit{Germanium} -- The germanium abundance $\log\epsilon$(Ge), derived from spectrum synthesis, is 3.7 $\pm$ 0.1. Relative to iron, Ge is overabundant with respect to solar by about a factor of 8 (0.9 dex). This provides the first measurement of germanium abundance in EHe stars.  

\textit{Strontium, Yr, Zr, and Ba} -- Relative to iron, the abundances of Sr, Yr, Zr, and Ba are found to be overabundant with respect to solar by about a factor of 20 (1.3 dex), 13 (1.1 dex), 16 (1.2 dex) and 8 (0.9 dex), respectively. A\,980 provides a maximum enhancement of $s$-process elements measured in a cool EHe star to date using significantly reliable number of transitions $-$ these are 4, 12, 16, and 4 for Sr, Yr, Zr, and Ba, respectively.
\section{Discussion and Conclusions}

A fine abundance analysis based on A\,980's observed high-resolution spectrum clearly demonstrates that it is a cool EHe star. 

The surface abundances of A\,980 are measured especially for the key elements: CNO, fluorine, and the heavier elements: Ge, Sr, Y, Zr, and Ba. The measured C/He ratio is about 0.2\% and has the lowest known C/He ratio. Note that the other EHes, excluding the four known EHes with a very low carbon abundance, exhibit a range in their C/He ratios: 0.3\% $-$ 1.3\%.

EHe stars are thought to be products of double white dwarf mergers. Case1 Predictions from Figure 5 of \cite{2019MNRAS.482.2320M} for CO+He white dwarf mergers are in broad agreement with the observed abundances of A\,980, in particular, the enhanced fluorine and the $s$-process elements (Y, Zr, Ba), compared to solar. 

A significant addition is the determination of germanium abundance in two cool EHes, A\,980 and LS IV -14\textdegree\ 109, with similar stellar parameters (T$_{\text{eff}}$, log\,g). These are the first determination of Ge abundances in EHe stars. The atmosphere of
A\,980 is significantly enhanced in germanium that is, eight times higher than the solar value relative to iron. This notable enhancement is evidence of the synthesis of germanium in a cool EHe star, A\,980. Note that A\,980 is enriched by about a factor of 4 in germanium (Ge/Fe) relative to LS IV -14\textdegree\ 109 while A\,980's enrichment in germanium is about a factor of 16 when compared with a normal subgiant HD\,107113 of similar metallicity; HD\,107113's Ge and Fe abundances are from \citet{2012ApJ...756...36R}. 

Two primary processes that could account for the synthesis of germanium are first the neutron-capture process (\citet{2007ApJ...656L..73K} and references therein) and second the rapid proton-capture process, known as the rp-process \citep{1994ApJ...420..364B}. Knowledge of the germanium production mechanism in A\,980 will provide new clues to the origin of EHe stars. To identify the process of germanium production, A\,980's observed abundances: Ge, light $s$-process (Sr, Y, Zr), and heavy $s$-process (Ba), were compared with \citet{2018MNRAS.477..421K} predictions for low- to intermediate-mass asymptotic giant branch (AGB) stars (see Table \ref{tab:ratiotable}). A\,980's mass is estimated to be 1 $\pm$ 0.5 M$_\odot$ based on its luminosity, that is calculated using the absolute magnitude from \citet{2022A&A...667A..83T}, along with our derived surface gravity (g) and effective temperature (T$_{\text{eff}}$). For AGB stars, as given in Table \ref{tab:ratiotable}, the abundance ratios are shaped by $s$-process nucleosynthesis. The expected ratios for low-mass AGB star (2.5M$_\odot$) are [Ge/Fe] = $+$0.5, [Ge/ls] = $-$1.1, and [ls/hs] = $-$0.7. However, for A\,980, the observed ratios are notably different: [Ge/Fe] = $+$0.87, [Ge/ls] = $-$0.34, and [ls/hs] = $+$0.36. Here, ls and hs represent the light and heavy $s$-process abundances, respectively. Specifically, the ratio [ls/hs] $>$ 0 in A\,980 contradicts low-mass AGB predictions where heavy $s$-process elements are expected to be more abundant. The expected ratios for intermediate-mass AGB star (5M$_\odot$), with different mass loss rates: VW93 and B95, are provided in Table 2 (see \citet{2018MNRAS.477..421K} for VW93 and B95). Here, the ratio [ls/hs] $>$ 0 for both cases and is in line with that observed for A\,980. The abundance ratios for a mass loss rate, VW93, are found to be in fair agreement with that observed for A\,980. This suggests that the atmospheres of A\,980 are contaminated by germanium produced via the $s$-process taking place in an intermediate-mass AGB star. Note that the abundance ratios depend on the adopted mass loss rate for intermediate-mass AGB stars but nucleosynthesis processes remain the same. 
\begin{deluxetable}{cccccc}
\tablecolumns{6}
\tablewidth{\hsize}
\tablecaption{Surface abundance ratios for comparative analysis.\label{tab:ratiotable}}
\tablehead{
\colhead{} & \colhead{A 980\tablenotemark{a}} & \colhead{CO+He\tablenotemark{b}} & \multicolumn{3}{c} {AGB\tablenotemark{c}} \\
\colhead{} & \colhead{} & \colhead{} & \colhead{2.5M$_\odot$} & \colhead{5M$_\odot$(VW93)} & \colhead{5M$_\odot$(B95)}
}
\startdata
$[$Fe$]$ & $-$0.80 & $-$1.55 & $-$0.70 & $-$0.70 & $-$0.70 \\
$[$Ge/Fe$]$ & $+$0.87 & ... & $+$0.50 & $+$0.90 & $+$0.30 \\
$[$Ge/ls$]$ & $-$0.34 & ... & $-$1.10 & $-$0.10 & $+$0.10  \\
$[$Ge/hs$]$ & $+$0.02 & ... & $-$1.80 & $+$0.50 & $+$0.25 \\
$[$ls/hs$]$ & $+$0.36 & $+$0.50 & $-$0.70 & $+$0.60 & $+$0.15 \\
$[$Sr/Fe$]$ & $+$1.33 & ...  & $+$1.50 & $+$1.10 & $+$0.25 \\
$[$Y/Fe$]$ & $+$1.13 & $+$1.35 & $+$1.60 & $+$1.05 & $+$0.20 \\
$[$Zr/Fe$]$ & $+$1.15 & $+$0.95 & $+$1.70 & $+$0.95 & $+$0.15 \\
$[$Ba/Fe$]$ & $+$0.85 & $+$0.65 & $+$2.30 & $+$0.40 & $+$0.05 \\
\enddata
\tablenotetext{a}{This work}
\tablenotetext{b}{Adopted from \citet{2019MNRAS.482.2320M}: Case1 Predictions}
\tablenotetext{c}{Adopted from \citet{2018MNRAS.477..421K}}
\end{deluxetable}

In comparison, the observed abundance ratios do suggest that the atmospheres of LS IV -14\textdegree\ 109 are not contaminated by freshly synthesized Ge and $s$-process elements (see Table \ref{tab:abundancetable}).

Alternatively, a significant enhancement in germanium abundance is evident from Thorne-\.Zytkow Objects (T\.ZOs) and presumably produce elements via the rapid proton-capture process (rp-process). T\.ZOs are proposed as the possible merged products of a neutron star (NS) and a non-degenerate star. \citet{2023MNRAS.524.1692F} provides the recent predictions of surface abundance ratios for Thorne-\.Zytkow Objects (T\.ZOs) but those are not in the regime of A\,980's metallicity and the helium fraction to perform a comparative analysis.

The abundance ratios [ls/Fe] and [hs/Fe] for A\,980 are similar to the predictions of CO+He WD mergers \citep{2019MNRAS.482.2320M}. However, predictions of germanium abundances are not available for CO+He white dwarf mergers to explore the origin of germanium in EHe stars. Theoretical studies for germanium synthesis in CO+He white dwarf mergers need to be explored.

\section{Acknowledgments} \label{sec:ackn}

We thank the referee for a constructive review. This research was supported by the Science and Engineering Research Board (SERB), DST, India through grant: CRG/2021/000108. Ajay expresses gratitude to Sriram Krishna for his assistance with installing and running SYNSPEC. He also thanks Fayaz S. and B.P. Hema for helping with the installation of the radiative transfer code EQWIDTH.

We thank the staff of IAO, Hanle, CREST, and Hosakote, who made these observations possible. The facilities at IAO and CREST are operated by the Indian Institute of Astrophysics, Bangalore.

\clearpage

\appendix
\restartappendixnumbering
\section{Error Analysis}\label{appedixerror}

The adopted stellar parameters are accurate to typically: $\Delta T_{\text{eff}}$ = $\pm$250 K, $\Delta$log \textit{g} = $\pm$0.3 cm s$^{-2}$, $\Delta \xi$ = $\pm$1.0 km s$^{-1}$. The line-to-line scatter, standard deviation due to the same ion holding many lines, and the uncertainty in stellar parameters are the major sources of errors in deriving abundances. The abundance errors due to line-to-line scatter are given in Table \ref{tab:linelist} while errors due to uncertainty in the stellar parameter are given in Table \ref{tab:errortable}. Typically, the error due to uncertainty in the stellar parameter was found to be lesser than the line-to-line scatter. The abundance derived from more lines is always better and more reliable than those derived from fewer lines. However, this has a negligible effect on the mean abundance of element E. Therefore, we also include a weight due to the number of lines that derive the ion's abundance. The mean abundance of an element E was calculated as -
$$\langle E\rangle = \frac{\langle E_a\rangle + \langle E_b\rangle}{2}$$
where $\langle E_a\rangle$ and $\langle E_b\rangle$ are mean abundances of an element E, calculated by giving weight due to line-to-line scatter and number of lines respectively. They are defined as - $$\langle E_a\rangle=\frac{w_{1a}\langle EI\rangle + w_{2a}\langle EII\rangle + ...}{w_{1a} + w_{2a}}$$
where $\langle EI\rangle, \langle EII\rangle,...$ are the abundances derived from neutrally-ionized, singly-ionized,... lines of element E. $w_{1a}, w_{2a},..$ are weights due to line-to-line scatter of ions ($\delta(EI),\delta(EII),...$), i.e., $$w_1=\frac{1}{(\delta(EI))^2}$$
And, $$\langle E_b\rangle=\frac{w_{1b}\langle EI\rangle + w_{2b}\langle EII\rangle + ...}{w_{1b} + w_{2b}}$$
where $w_{1b}, w_{2b},..$ are weights due to the number of lines used to derive the abundance of ions, i.e., $$w_{1b} = \text{\# of lines used to derive } \langle EI\rangle$$
\startlongtable
\begin{deluxetable}{lcccl}
\tablecolumns{5}
\tablewidth{\hsize} 
\tablecaption{Errors in elemental abundances due to uncertainties in the stellar parameters. The abundance error due to $\Delta\text{T}_{\text{eff}}$ is the difference in abundances derived from the adopted model (T$_{\text{eff}}$, log\,g, $\xi$) and a model (T$_{\text{eff}}$+$\Delta\text{T}_{\text{eff}}$, log\,g, $\xi$). The abundance error due to $\Delta$log\,g is the difference in abundances derived from the adopted model (T$_{\text{eff}}$, log\,g, $\xi$) and a model (T$_{\text{eff}}$, log\,g+$\Delta$log\,g, $\xi$). The abundance error due to $\Delta\xi$ is the difference in the abundances derived from the adopted model (T$_{\text{eff}}$, log\,g, $\xi$) and a model (T$_{\text{eff}}$, log\,g, $\xi$+$\Delta\xi$).\label{tab:errortable}}
\tablehead{
\colhead{Species} & \colhead{$\Delta\text{T}_{\text{eff}}=+250$} & \colhead{$\Delta$log\,g$=+0.25$} & \colhead{$\Delta\xi=+1.0$} & \colhead{rms}\\
\colhead{} & \colhead{(K)} & \colhead{(cgs)} & \colhead{(km s$^{-1}$)} & \colhead{}
}
\startdata
C I & $+$0.12 & $-$0.06 & $-$0.03 & 0.13\\
C II & $-$0.17 & $+$0.08 & $-$0.15 & 0.24\\
N I & $+$0.08 & $-$0.03 & $-$0.02 & 0.09\\
N II & $-$0.12 & $+$0.06 & $-$0.03 & 0.14\\
O I & $+$0.06 & $-$0.02 & $-$0.05 & 0.08\\
Ne I & $-$0.16 & $+$0.10 & $-$0.09 & 0.21\\
Na I & $+$0.12 & $-$0.06 & $-$0.04 & 0.14\\
Mg I & $+$0.23 & $-$0.10 & $-$0.03 & 0.25\\
Mg II & $+$0.03 & 0.00 & $-$0.01 & 0.03\\
Al II & $-$0.10 & $+$0.07 & $-$0.08 & 0.15\\
Si I & $+$0.15 & $-$0.06 & 0.00 & 0.16\\
Si II & $-$0.08 & $+$0.06 & $-$0.05 & 0.11\\
P II & $-$0.09 & $+$0.09 & $-$0.03 & 0.13\\
S I & $+$0.13 & $-$0.06 & $-$0.01 & 0.14\\
S II & $-$0.09 & $+$0.07 & $-$0.05 & 0.13\\
Ca II & $+$0.14 & $-$0.05 & $-$0.02 & 0.15\\
Sc II & $+$0.23 & $-$0.03 & $-$0.01 & 0.23\\
Ti II & $+$0.17 & $-$0.01 & $-$0.04 & 0.17\\
V II & $+$0.14 & $+$0.02 & $-$0.02 & 0.14\\
Cr II & $+$0.08 & $+$0.03 & $-$0.04 & 0.09\\
Mn II & $+$0.06 & $+$0.04 & $-$0.02 & 0.07\\
Fe I & $+$0.21 & $-$0.08 & $-$0.01 & 0.22\\
Fe II & $-$0.01 & $+$0.04 & $-$0.03 & 0.05\\
Co II & $+$0.04 & $+$0.05 & $-$0.03 & 0.07\\
Ni I & $+$0.17 & $-$0.07 & 0.00 & 0.18\\
Ni II & 0.00 & $+$0.05 & $-$0.03 & 0.06\\
Ge II & $+$0.05 & $+$0.12 & $+$0.04 & 0.14\\
Sr II & $+$0.25 & $-$0.05 & $-$0.18 & 0.30\\
Y II & $+$0.26 & $-$0.04 & $-$0.03 & 0.27\\
Zr II & $+$0.21 & $-$0.02 & $-$0.03 & 0.22\\
Ba II & $+$0.26 & $-$0.06 & $-$0.05 & 0.27\\
\enddata
\end{deluxetable}
\section{LINES USED FOR ABUNDANCE ANALYSIS}\label{linelist}

The lines used for the abundance analysis of A980 are given in Table \ref{tab:linelist}. Also, the lower excitation potential ($\chi$), log \textit{gf}, equivalent width (W$_{\lambda}$), and abundance (log $\epsilon$) derived for each line are listed. The elemental abundances are derived by using model atmospheres with T$_{\text{eff}}$ = 9250 K, log g = 1.25 cm s$^{-2}$, C/He = 0.3\% and microturbulent velocity $\xi$ = 11.5 km s$^{-1}$.

\startlongtable
\begin{deluxetable}{lccccl}
\tablecolumns{6}
\tablewidth{\hsize} 
\tablecaption{Lines used to derive elemental abundances for A980. \label{tab:linelist}}
\tablehead{
\colhead{Ion} & \colhead{} & \colhead{} & \colhead{W$_{\lambda}$} & \colhead{} & \colhead{} \\
\colhead{$\lambda$ (\AA)} & \colhead{$\chi$ (eV)} & \colhead{log \textit{gf}} & \colhead{(m\AA)} & \colhead{log $\epsilon$\tablenotemark{a}} &\colhead{Reference\tablenotemark{b}}
}
\startdata
H I &   &   &   &   &   \\
6562.82 & 10.20 & 0.710 & Synth & 5.62 & NIST\\
4861.33 & 10.20 & -0.020 & Synth & 5.80 & NIST\\
4340.47 & 10.20 & -0.447 & Synth & 6.00 & NIST\\
  &   &   &   &   &   \\
Mean: &   &   &   & 5.81 $\pm$ 0.16 &   \\
  &   &   &   &   &   \\
He I &   &   &   &   &   \\
5048.00 & 21.22 & -1.587 & Synth & 11.54 & NIST \\
  &   &   &   &   &   \\
C I &   &   &   &   &   \\
6595.24 & 8.85 & -2.400 & 104 & 8.84 & NIST \\
6591.45 & 8.85 & -2.400 & 101 & 8.82 & NIST \\
6683.95 & 8.85 & -2.230 & 216 & 9.16 & NIST \\
6688.79 & 8.85 & -2.130 & 188 & 8.95 & NIST \\
6711.29 & 8.86 & -2.690 & 95 & 9.09 & NIST \\
7022.24 & 8.64 & -2.670 & 114 & 9.05 & NIST \\
7074.86 & 8.64 & -2.120 & 159 & 8.71 & NIST \\
7132.11 & 8.65 & -2.200 & 127 & 8.66 & NIST \\
7473.31 & 8.77 & -2.040 & 200 & 8.89 & NIST \\
7837.11 & 8.85 & -1.780 & 193 & 8.63 & NIST \\
7840.27 & 8.85 & -1.840 & 191 & 8.68 & NIST \\
7852.86 & 8.85 & -1.680 & 205 & 8.58 & NIST \\
8018.56 & 8.85 & -2.100 & 113 & 8.58 & NIST \\
8078.48 & 8.85 & -1.820 & 208 & 8.70 & NIST \\
5824.64 & 8.85 & -2.600 & 127 & 9.13 & NIST \\
6012.24 & 8.64 & -2.000 & 220 & 8.79 & NIST \\
6078.40 & 8.85 & -2.300 & 167 & 9.01 & NIST \\
6108.53 & 8.85 & -2.500 & 113 & 8.96 & NIST \\
5540.76 & 8.64 & -2.400 & 200 & 9.12 & NIST \\
5547.27 & 8.64 & -2.300 & 201 & 9.02 & NIST \\
4817.37 & 7.48 & -3.040 & 210 & 9.16 & NIST \\
4943.58 & 8.65 & -2.460 & 159 & 9.03 & NIST \\
5159.92 & 8.64 & -2.150 & 175 & 8.77 & NIST \\
4355.41 & 7.68 & -3.330 & 80 & 9.01 & NIST \\
6671.84 & 8.85 & -1.650 & 273 & 8.79 & NIST \\
7087.83 & 8.65 & -1.440 & 279 & 8.48 & NIST \\
7093.25 & 8.65 & -1.700 & 270 & 8.72 & NIST \\
7685.17 & 8.77 & -1.520 & 294 & 8.70 & NIST \\
7832.63 & 8.85 & -1.810 & 277 & 8.97 & NIST \\
8014.98 & 8.85 & -1.600 & 280 & 8.75 & NIST \\
8062.36 & 8.85 & -1.860 & 255 & 8.91 & NIST \\
8070.42 & 8.85 & -1.950 & 266 & 9.04 & NIST \\
8083.80 & 8.85 & -1.730 & 248 & 8.75 & NIST \\
5793.12 & 7.95 & -2.060 & 295 & 8.71 & NIST \\
6001.13 & 8.64 & -2.100 & 297 & 9.17 & NIST \\
6007.18 & 8.64 & -2.100 & 224 & 8.91 & NIST \\
6010.68 & 8.64 & -1.900 & 257 & 8.83 & NIST \\
6016.45 & 8.64 & -1.800 & 245 & 8.68 & NIST \\
5551.03 & 8.65 & -1.600 & 280 & 8.61 & NIST \\
5548.90 & 8.65 & -1.760 & 271 & 8.74 & KP \\
4734.26 & 7.95 & -2.370 & 279 & 9.02 & NIST \\
4766.67 & 7.48 & -2.620 & 255 & 8.91 & NIST \\
7202.26 & 9.00 & -1.900 & 199 & 8.87 & NIST \\
7216.03 & 9.17 & -2.300 & 139 & 9.12 & NIST \\
7364.73 & 9.00 & -1.840 & 216 & 8.88 & NIST \\
7987.89 & 9.17 & -2.100 & 169 & 9.03 & NIST \\
  &   &   &   &   &   \\
Mean: &   &   &   & 8.87 $\pm$ 0.18 &   \\
  &   &   &   &   &   \\
C II &   &   &   &   &   \\
6578.05 & 14.45 & -0.022 & 610 & 9.61 & NIST \\
6582.88 & 14.45 & -0.323 & 515 & 9.46 & NIST \\
7231.32 & 16.33 & 0.038 & 360 & 9.51 & NIST \\
  &   &   &   &   &   \\
Mean: &   &   &   & 9.53 $\pm$ 0.07 &   \\
  &   &   &   &   &   \\
N I &   &   &   &   &   \\
6622.53 & 11.76 & -1.504 & 215 & 8.53 & NIST \\
6644.96 & 11.76 & -0.858 & 270 & 8.09 & NIST \\
6653.41 & 11.76 & -1.138 & 220 & 8.19 & NIST \\
6637.01 & 11.75 & -1.432 & 186 & 8.34 & NIST \\
6646.52 & 11.75 & -1.539 & 134 & 8.22 & NIST \\
6708.76 & 11.84 & -1.288 & 210 & 8.35 & NIST \\
6926.67 & 11.84 & -1.475 & 131 & 8.20 & NIST \\
6945.18 & 11.84 & -1.100 & 241 & 8.30 & NIST \\
6951.60 & 11.84 & -1.525 & 204 & 8.57 & NIST \\
6979.19 & 11.84 & -1.542 & 118 & 8.20 & NIST \\
6982.03 & 11.84 & -1.525 & 174 & 8.45 & NIST \\
7485.18 & 12.01 & -1.572 & 135 & 8.44 & NIST \\
7546.21 & 12.00 & -1.313 & 150 & 8.24 & NIST \\
7550.91 & 12.01 & -1.213 & 172 & 8.25 & NIST \\
5616.56 & 11.76 & -1.370 & 246 & 8.48 & KP \\
5600.53 & 11.76 & -2.020 & 118 & 8.58 & KP \\
5816.49 & 11.79 & -2.070 & 108 & 8.60 & NBS \\
5856.00 & 11.79 & -2.110 & 80 & 8.47 & NBS \\
6008.47 & 11.60 & -1.406 & 215 & 8.31 & NIST \\
5378.27 & 10.93 & -2.841 & 50 & 8.50 & NIST \\
5557.38 & 11.75 & -1.735 & 83 & 8.09 & NIST \\
5560.34 & 11.75 & -1.181 & 192 & 8.07 & NIST \\
4935.12 & 10.69 & -1.891 & 165 & 8.09 & NIST \\
4963.99 & 11.76 & -2.140 & 111 & 8.67 & KP \\
6471.03 & 11.75 & -1.890 & 80 & 8.26 & KP \\
  &   &   &   &   &   \\
Mean: &   &   &   & 8.34 $\pm$ 0.18 &   \\
  &   &   &   &   &   \\
N II &   &   &   &   &   \\
4643.08 & 18.48 & -0.371 & 63 & 8.34 & NIST \\
4607.16 & 18.46 & -0.522 & 55 & 8.38 & NIST \\
5679.56 & 18.48 & 0.225 & 130 & 8.63 & NIST \\
5686.21 & 18.47 & -0.586 & 65 & 8.78 & NIST \\
  &   &   &   &   &   \\
Mean: &   &   &   & 8.53 $\pm$ 0.21 &   \\
  &   &   &   &   &   \\
O I &   &   &   &   &   \\
6453.60 & 10.74 & -1.288 & 224 & 7.95 & NIST \\
6454.44 & 10.74 & -1.066 & 220 & 7.72 & NIST \\
7156.70 & 12.73 & 0.288 & 268 & 7.74 & NIST \\
7950.83 & 12.54 & 0.340 & 382 & 8.03 & NIST \\
7952.16 & 12.54 & 0.170 & 342 & 8.04 & NIST \\
7943.15 & 12.54 & -0.550 & 107 & 7.73 & NIST \\
6155.98 & 10.69 & -0.660 & 346 & 7.75 & Luck \\
6156.77 & 10.69 & -0.440 & 399 & 7.74 & NIST \\
5330.66 & 10.69 & -0.970 & 269 & 7.74 & NIST \\
5435.78 & 10.74 & -1.544 & 210 & 8.12 & NIST \\
5436.86 & 10.74 & -1.399 & 265 & 8.18 & NIST \\
  &   &   &   &   &   \\
Mean: &   &   &   & 7.89 $\pm$ 0.18 &   \\
  &   &   &   &   &   \\
F I &   &   &   &   &   \\
6902.48 & 12.73 & 0.180 & Synth & 6.00 & NIST \\
6909.82 & 12.75 & -0.230 & Synth & 6.35 & NIST \\
6834.26 & 12.73 & -0.210 & Synth & 6.13 & NIST \\
7331.96 & 12.70 & -0.110 & Synth & 6.06 & NIST \\
7398.69 & 12.70 & 0.240 & Synth & 6.00 & NIST \\
7754.70 & 12.98 & 0.240 & Synth & 6.06 & NIST \\
  &   &   &   &   &   \\
Mean: &   &   &   & 6.10 $\pm$ 0.12 &   \\
  &   &   &   &   &   \\
Ne I &   &   &   &   &   \\
7032.41 & 16.62 & -0.228 & 286 & 8.71 & NIST \\
6334.43 & 16.62 & -0.310 & 321 & 8.82 & NIST \\
6217.28 & 16.62 & -0.960 & 207 & 8.68 & NIST \\
6143.06 & 16.62 & -0.100 & 336 & 8.64 & NIST \\
5975.53 & 16.62 & -1.250 & 145 & 8.46 & NIST \\
5944.83 & 16.62 & -0.520 & 249 & 8.44 & NIST \\
5881.90 & 16.62 & -0.750 & 215 & 8.43 & NIST \\
7245.17 & 16.67 & -0.622 & 269 & 9.08 & NIST \\
6506.53 & 16.67 & -0.020 & 326 & 8.64 & NIST \\
6074.34 & 16.67 & -0.480 & 245 & 8.44 & NIST \\
6382.99 & 16.67 & -0.230 & 283 & 8.54 & NIST \\
6929.47 & 16.85 & -0.200 & 271 & 8.69 & NIST \\
7024.05 & 16.85 & -1.380 & 72 & 8.34 & NIST \\
6532.88 & 16.72 & -0.680 & 212 & 8.59 & NIST \\
6717.04 & 16.85 & -0.360 & 294 & 8.94 & NIST \\
6266.50 & 16.72 & -0.360 & 310 & 8.83 & NIST \\
5852.49 & 16.85 & -0.500 & 260 & 8.59 & NIST \\
6163.59 & 16.72 & -0.600 & 246 & 8.62 & NIST \\
7438.90 & 16.72 & -1.211 & 137 & 8.80 & NIST \\
7488.87 & 18.38 & -0.010 & 156 & 8.62 & NIST \\
7535.77 & 18.38 & -0.110 & 176 & 8.89 & NIST \\
7544.04 & 18.38 & -0.480 & 116 & 8.76 & NIST \\
8377.61 & 18.56 & 0.670 & 271 & 8.67 & NIST \\
6074.34 & 16.67 & -0.480 & 245 & 8.44 & NIST \\
  &   &   &   &   &   \\
Mean: &   &   &   & 8.65 $\pm$ 0.18 &   \\
  &   &   &   &   &   \\
Na I &   &   &   &   &   \\
5682.63 & 2.10 & -0.706 & 179 & 6.33 & NIST \\
5688.21 & 2.10 & -0.452 & 257 & 6.36 & NIST \\
6154.23 & 2.10 & -1.547 & 74 & 6.65 & NIST \\
4664.81 & 2.10 & -1.565 & 70 & 6.67 & NIST \\
8194.82 & 2.10 & 0.492 & 530 & 6.28 & NIST \\
  &   &   &   &   &   \\
Mean: &   &   &   & 6.46 $\pm$ 0.19 &   \\
  &   &   &   &   &   \\
Mg I &   &   &   &   &   \\
4702.99 & 4.35 & -0.440 & 213 & 6.69 & NIST \\
5528.41 & 4.35 & -0.498 & 248 & 6.79 & NIST \\
  &   &   &   &   &   \\
Mean: &   &   &   & 6.74 $\pm$ 0.08 &   \\
  &   &   &   &   &   \\
Mg II &   &   &   &   &   \\
5923.37 & 12.08 & -1.532 & 97 & 7.14 & NIST \\
5943.50 & 12.08 & -1.542 & 145 & 7.42 & NIST \\
5464.14 & 12.08 & -1.702 & 81 & 7.20 & NIST \\
5460.02 & 12.08 & -2.002 & 70 & 7.42 & NIST \\
  &   &   &   &   &   \\
Mean: &   &   &   & 7.29 $\pm$ 0.14 &   \\
  &   &   &   &   &   \\
Al II &   &   &   &   &   \\
6823.48 & 13.07 & -0.123 & 191 & 6.13 & NIST \\
6837.14 & 13.08 & 0.097 & 217 & 6.07 & NIST \\
7063.64 & 11.32 & -0.368 & 372 & 6.39 & NIST \\
6231.78 & 13.07 & 0.389 & 287 & 6.04 & NIST \\
5593.30 & 13.26 & 0.337 & 272 & 6.02 & NIST \\
6920.34 & 13.26 & -0.144 & 150 & 6.03 & NIST \\
  &   &   &   &   &   \\
Mean: &   &   &   & 6.12 $\pm$ 0.14 &   \\
  &   &   &   &   &   \\
Al III &   &   &   &   &   \\
5696.60 & 15.64 & 0.232 & 84 & 6.05 & NIST \\
5722.73 & 15.64 & -0.071 & 80 & 6.32 & NIST \\
  &   &   &   &   &   \\
Mean: &   &   &   & 6.18 $\pm$ 0.13 &   \\
  &   &   &   &   &   \\
Si I &   &   &   &   &   \\
7034.90 & 5.87 & -0.880 & 63 & 7.19 & GARZ \\
7932.20 & 5.96 & -0.470 & 121 & 7.17 & NIST \\
5708.44 & 4.93 & -1.470 & 38 & 6.98 & NIST \\
6145.08 & 5.61 & -1.480 & 70 & 7.67 & Luck \\
6155.22 & 5.62 & -0.750 & 115 & 7.21 & Tom97 \\
  &   &   &   &   &   \\
Mean: &   &   &   & 7.24 $\pm$ 0.26 &   \\
  &   &   &   &   &   \\
Si II &   &   &   &   &   \\
6660.52 & 14.50 & 0.162 & 207 & 7.46 & NIST \\
6665.00 & 14.49 & -0.240 & 160 & 7.56 & NIST \\
6671.88 & 14.53 & 0.409 & 282 & 7.67 & NIST \\
5868.40 & 14.53 & 0.400 & 298 & 7.58 & Kurucz \\
4621.42 & 12.52 & -0.608 & 245 & 7.14 & NIST \\
5706.37 & 14.17 & -0.225 & 118 & 6.93 & NIST \\
5632.97 & 14.19 & -0.818 & 68 & 7.14 & NIST \\
  &   &   &   &   &   \\
Mean: &   &   &   & 7.35 $\pm$ 0.28 &   \\
  &   &   &   &   &   \\
P II &   &   &   &   &   \\
6034.01 & 10.74 & -0.220 & 136 & 5.40 & NIST \\
6043.10 & 10.80 & 0.420 & 209 & 5.23 & NIST \\
6503.46 & 10.91 & -0.006 & 118 & 5.26 & NIST \\
5344.75 & 10.74 & -0.390 & 154 & 5.59 & NIST \\
5499.73 & 10.80 & -0.300 & 96 & 5.17 & NIST \\
7845.63 & 11.02 & -0.040 & 119 & 5.51 & NIST \\
  &   &   &   &   &   \\
Mean: &   &   &   & 5.36 $\pm$ 0.17 &   \\
  &   &   &   &   &   \\
S I &   &   &   &   &   \\
6748.79 & 7.87 & -0.638 & 140 & 7.15 & NIST \\
6757.16 & 7.87 & -0.351 & 228 & 7.20 & NIST \\
6052.66 & 7.87 & -0.672 & 109 & 7.01 & NIST \\
4694.11 & 6.52 & -1.713 & 44 & 6.86 & NIST \\
4695.44 & 6.52 & -1.871 & 42 & 6.99 & NIST \\
  &   &   &   &   &   \\
Mean: &   &   &   & 7.04 $\pm$ 0.14 &   \\
  &   &   &   &   &   \\
S II &   &   &   &   &   \\
5606.11 & 13.73 & 0.124 & 252 & 6.97 & NIST \\
5664.73 & 13.66 & -0.427 & 115 & 6.61 & NIST \\
5428.67 & 13.58 & -0.177 & 202 & 6.86 & NIST \\
5453.83 & 13.67 & 0.442 & 300 & 6.88 & NIST \\
5556.01 & 13.62 & -1.020 & 64 & 6.73 & NIST \\
4716.27 & 13.62 & -0.365 & 163 & 6.74 & NIST \\
4885.65 & 14.00 & -0.674 & 120 & 6.94 & NIST \\
4917.21 & 14.00 & -0.375 & 130 & 6.72 & NIST \\
4925.35 & 13.58 & -0.206 & 220 & 6.93 & NIST \\
4486.64 & 15.87 & -0.400 & 40 & 6.74 & NIST \\
  &   &   &   &   &   \\
Mean: &   &   &   & 6.81 $\pm$ 0.12 &   \\
  &   &   &   &   &   \\
Ar I &   &   &   &   &   \\
7067.22 & 11.55 & -0.850 & 70 & 6.21 & NIST \\
7383.98 & 11.62 & -0.460 & 159 & 6.29 & NIST \\
8264.52 & 11.83 & -0.328 & 170 & 6.20 & NIST \\
8006.16 & 11.62 & -0.630 & 71 & 5.98 & NIST \\
8103.69 & 11.62 & -0.131 & 160 & 5.91 & NIST \\
8408.21 & 11.83 & 0.073 & 181 & 5.82 & NIST \\
8115.31 & 11.55 & 0.360 & 280 & 5.76 & NIST \\
  &   &   &   &   &   \\
Mean: &   &   &   & 5.99 $\pm$ 0.19 &   \\
  &   &   &   &   &   \\
Ca I &   &   &   &   &   \\
4226.73 & 0.00 & 0.244 & 212 & 5.55 & NIST \\
  &   &   &   &   &   \\
Ca II &   &   &   &   &   \\
8254.73 & 7.51 & -0.390 & 296 & 5.61 & NIST \\
5285.27 & 7.50 & -1.180 & 122 & 5.83 & NIST \\
4097.10 & 7.50 & -1.000 & 127 & 5.76 & NIST \\
  &   &   &   &   &   \\
Mean: &   &   &   & 5.73 $\pm$ 0.11 &   \\
  &   &   &   &   &   \\
Sc II &   &   &   &   &   \\
6309.90 & 1.50 & -1.570 & 140 & 3.08 & NIST \\
6279.76 & 1.50 & -1.210 & 232 & 3.03 & NIST \\
6245.63 & 1.51 & -0.980 & 236 & 2.81 & NIST \\
6604.60 & 1.36 & -1.310 & 202 & 2.93 & NIST \\
5684.19 & 1.51 & -1.070 & 271 & 3.04 & NIST \\
4420.66 & 0.62 & -2.270 & 108 & 3.23 & NIST \\
4014.48 & 0.32 & -1.660 & 165 & 2.73 & NIST \\
  &   &   &   &   &   \\
Mean: &   &   &   & 2.98 $\pm$ 0.17 &   \\
  &   &   &   &   &   \\
Ti II &   &   &   &   &   \\
6559.58 & 2.05 & -2.019 & 147 & 3.99 & NIST \\
6491.61 & 2.06 & -1.793 & 314 & 4.28 & NIST \\
6607.02 & 2.06 & -2.790 & 49 & 4.21 & NIST \\
7355.45 & 2.60 & -1.916 & 210 & 4.44 & NIST \\
7214.73 & 2.59 & -1.750 & 206 & 4.25 & NIST \\
5381.02 & 1.57 & -1.921 & 316 & 4.17 & NIST \\
5418.77 & 1.58 & -2.002 & 230 & 4.00 & NIST \\
4779.98 & 2.05 & -1.370 & 345 & 4.10 & NIST \\
4798.53 & 1.08 & -2.679 & 253 & 4.50 & NIST \\
4865.61 & 1.16 & -2.788 & 237 & 4.60 & NIST \\
4874.01 & 3.09 & -0.805 & 322 & 4.11 & NIST \\
4911.20 & 3.12 & -0.609 & 375 & 4.10 & NIST \\
5010.21 & 3.09 & -1.291 & 275 & 4.43 & NIST \\
4227.33 & 1.13 & -2.236 & 352 & 4.51 & NIST \\
4287.87 & 1.08 & -2.020 & 398 & 4.42 & NIST \\
4316.79 & 2.05 & -1.577 & 281 & 4.18 & NIST \\
4421.94 & 2.06 & -1.663 & 317 & 4.38 & NIST \\
4350.84 & 2.06 & -1.735 & 245 & 4.22 & NIST \\
4367.65 & 2.59 & -0.862 & 399 & 4.23 & NIST \\
4395.84 & 1.24 & -1.928 & 316 & 4.12 & NIST \\
4394.06 & 1.22 & -1.784 & 342 & 4.05 & NIST \\
4411.07 & 3.09 & -0.667 & 348 & 4.14 & NIST \\
4518.33 & 1.08 & -2.555 & 220 & 4.31 & NIST \\
4524.68 & 1.23 & -2.343 & 236 & 4.25 & NIST \\
4544.02 & 1.24 & -2.410 & 200 & 4.20 & NIST \\
4568.32 & 1.22 & -2.650 & 165 & 4.30 & NIST \\
4529.48 & 1.57 & -1.638 & 374 & 4.21 & NIST \\
4609.27 & 1.18 & -3.260 & 110 & 4.65 & NIST \\
4636.32 & 1.16 & -2.855 & 95 & 4.15 & NIST \\
6513.04 & 4.00 & -1.310 & 113 & 4.36 & NIST \\
4158.28 & 5.43 & -0.475 & 100 & 4.46 & K88 \\
  &   &   &   &   &   \\
Mean: &   &   &   & 4.27 $\pm$ 0.17 &   \\
  &   &   &   &   &   \\
V II &   &   &   &   &   \\
4528.48 & 2.28 & -1.050 & 263 & 3.47 & NIST \\
4564.58 & 2.27 & -1.210 & 213 & 3.46 & NIST \\
4600.17 & 2.26 & -1.360 & 173 & 3.45 & NIST \\
3977.72 & 1.48 & -1.570 & 170 & 3.26 & NIST \\
4036.76 & 1.48 & -1.570 & 200 & 3.36 & NIST \\
  &   &   &   &   &   \\
Mean: &   &   &   & 3.40 $\pm$ 0.09 &   \\
  &   &   &   &   &   \\
Cr II &   &   &   &   &   \\
6053.47 & 4.74 & -2.150 & 193 & 4.88 & NIST \\
5246.75 & 3.71 & -2.460 & 210 & 4.66 & NIST \\
5249.40 & 3.74 & -2.620 & 223 & 4.88 & NIST \\
5420.90 & 3.76 & -2.360 & 211 & 4.59 & NIST \\
5407.62 & 3.81 & -2.088 & 265 & 4.53 & NIST \\
5308.42 & 4.07 & -1.810 & 359 & 4.73 & NIST \\
5310.70 & 4.07 & -2.270 & 296 & 4.97 & NIST \\
5334.88 & 4.05 & -1.562 & 383 & 4.55 & NIST \\
5478.35 & 4.16 & -1.908 & 369 & 4.91 & NIST \\
5502.05 & 4.15 & -1.990 & 311 & 4.78 & NIST \\
5508.60 & 4.14 & -2.120 & 330 & 4.97 & NIST \\
4207.35 & 3.81 & -2.457 & 220 & 4.87 & NIST \\
4275.57 & 3.86 & -1.709 & 358 & 4.66 & NIST \\
4284.19 & 3.85 & -1.864 & 351 & 4.78 & NIST \\
4539.59 & 4.04 & -2.530 & 243 & 5.11 & NIST \\
4565.74 & 4.04 & -2.110 & 271 & 4.79 & NIST \\
4112.59 & 3.09 & -3.019 & 123 & 4.63 & NIST \\
4113.24 & 3.09 & -2.274 & 234 & 4.32 & NIST \\
5620.63 & 6.46 & -1.145 & 178 & 4.84 & K88 \\
5678.42 & 6.46 & -1.238 & 119 & 4.68 & K88 \\
4901.62 & 6.49 & -0.826 & 178 & 4.57 & K88 \\
4912.45 & 6.48 & -0.948 & 212 & 4.81 & K88 \\
4145.77 & 5.32 & -1.164 & 324 & 4.88 & K88 \\
4098.44 & 5.31 & -1.470 & 165 & 4.57 & K88 \\
  &   &   &   &   &   \\
Mean: &   &   &   & 4.75 $\pm$ 0.18 &   \\
  &   &   &   &   &   \\
Mn II &   &   &   &   &   \\
7252.40 & 10.77 & 0.680 & 85 & 4.91 & NIST \\
7330.58 & 3.71 & -2.713 & 212 & 4.80 & K88\\
7347.81 & 3.71 & -3.184 & 183 & 5.18 & K88\\
7353.55 & 3.70 & -2.726 & 258 & 4.96 & K88\\
7415.80 & 3.71 & -2.202 & 343 & 4.72 & NIST \\
7432.29 & 3.71 & -2.498 & 221 & 4.63 & NIST \\
6122.43 & 10.18 & 0.950 & 225 & 4.94 & NIST \\
5295.40 & 9.86 & 0.360 & 217 & 5.28 & NIST \\
5296.97 & 9.86 & 0.620 & 237 & 5.11 & NIST \\
5302.32 & 9.86 & 1.000 & 313 & 5.06 & NIST \\
5559.05 & 6.18 & -1.318 & 151 & 4.64 & NIST \\
4727.86 & 5.37 & -2.017 & 139 & 4.86 & K88 \\
4755.73 & 5.40 & -1.242 & 300 & 4.71 & K88 \\
4730.38 & 5.37 & -2.147 & 132 & 4.96 & NIST \\
4206.37 & 5.40 & -1.540 & 189 & 4.67 & NIST \\
4259.19 & 5.40 & -1.440 & 233 & 4.72 & NIST \\
4238.78 & 1.83 & -3.626 & 248 & 4.85 & NIST \\
4377.74 & 5.44 & -2.144 & 61 & 4.61 & NIST \\
4105.00 & 6.13 & -1.349 & 170 & 4.82 & NIST \\
4136.94 & 6.14 & -1.290 & 151 & 4.67 & NIST \\
  &   &   &   &   &   \\
Mean: &   &   &   & 4.85 $\pm$ 0.19 &   \\
  &   &   &   &   &   \\
Fe I &   &   &   &   &   \\
5383.37 & 4.31 & 0.645 & 224 & 6.73 & NIST \\
5424.07 & 4.32 & 0.520 & 176 & 6.69 & FMW \\
5615.64 & 3.33 & 0.050 & 135 & 6.40 & NIST \\
4891.49 & 2.85 & -0.112 & 136 & 6.32 & NIST \\
4903.31 & 2.88 & -0.926 & 55 & 6.67 & NIST \\
4918.99 & 2.86 & -0.342 & 149 & 6.61 & NIST \\
5162.29 & 4.16 & 0.020 & 82 & 6.67 & NIST \\
4920.50 & 2.83 & 0.068 & 338 & 6.83 & NIST \\
  &   &   &   &   &   \\
Mean: &   &   &   & 6.61 $\pm$ 0.17 &   \\
  &   &   &   &   &   \\
Fe II &   &   &   &   &   \\
6386.75 & 6.77 & -2.600 & 140 & 6.78 & NIST \\
6487.43 & 6.78 & -2.500 & 156 & 6.77 & NIST \\
6770.90 & 11.20 & -0.360 & 120 & 6.89 & NIST \\
6792.54 & 11.22 & -0.110 & 98 & 6.52 & NIST \\
6857.85 & 11.22 & 0.610 & 234 & 6.53 & NIST \\
6932.00 & 11.31 & 0.400 & 175 & 6.52 & NIST \\
6927.88 & 11.26 & 0.670 & 235 & 6.51 & NIST \\
7301.56 & 3.89 & -4.000 & 185 & 6.74 & NIST \\
7417.54 & 5.82 & -3.486 & 84 & 6.87 & NIST \\
7496.69 & 11.08 & -0.610 & 43 & 6.53 & NIST \\
7506.54 & 9.74 & -0.420 & 201 & 6.63 & NIST \\
7690.51 & 9.78 & -0.260 & 241 & 6.68 & NIST \\
7780.37 & 9.76 & -0.400 & 218 & 6.70 & NIST \\
7801.25 & 5.90 & -2.900 & 143 & 6.64 & NIST \\
7548.66 & 11.30 & 0.090 & 137 & 6.67 & NIST \\
7543.38 & 11.35 & -0.540 & 75 & 6.91 & NIST \\
7688.45 & 11.08 & -0.070 & 149 & 6.77 & NIST \\
7748.36 & 11.05 & -0.090 & 161 & 6.84 & NIST \\
7739.16 & 11.11 & -0.900 & 21 & 6.49 & NIST \\
7969.04 & 11.35 & -0.640 & 89 & 7.10 & NIST \\
7718.51 & 11.30 & -0.180 & 124 & 6.86 & NIST \\
7829.73 & 7.52 & -2.130 & 110 & 6.63 & NIST \\
7851.94 & 6.23 & -2.700 & 160 & 6.71 & NIST \\
7866.56 & 5.55 & -3.321 & 89 & 6.57 & NIST \\
7981.91 & 9.65 & -0.560 & 180 & 6.60 & NIST \\
5627.49 & 3.39 & -4.080 & 233 & 6.68 & NIST \\
5725.95 & 3.42 & -4.800 & 157 & 7.14 & NIST \\
5907.37 & 7.81 & -2.300 & 110 & 6.88 & NIST \\
5648.90 & 10.56 & -0.170 & 240 & 6.85 & NIST \\
5726.55 & 10.71 & -0.040 & 177 & 6.52 & NIST \\
5746.57 & 10.63 & -0.400 & 186 & 6.87 & NIST \\
5751.49 & 10.63 & -0.610 & 115 & 6.72 & NIST \\
5838.99 & 10.84 & -0.600 & 106 & 6.77 & NIST \\
5871.77 & 10.83 & -0.280 & 191 & 6.88 & NIST \\
5842.29 & 10.74 & -0.330 & 201 & 6.93 & NIST \\
5891.36 & 7.24 & -2.030 & 247 & 6.90 & NIST \\
5948.42 & 10.74 & -0.200 & 214 & 6.87 & NIST \\
5965.62 & 10.68 & 0.070 & 244 & 6.71 & NIST \\
5976.68 & 10.68 & -0.330 & 164 & 6.74 & NIST \\
5981.75 & 7.87 & -2.200 & 80 & 6.64 & NIST \\
6060.97 & 7.80 & -1.690 & 200 & 6.69 & NIST \\
6049.44 & 10.71 & -0.370 & 159 & 6.78 & NIST \\
6229.34 & 2.82 & -4.824 & 121 & 6.64 & NIST \\
5241.05 & 10.39 & -0.580 & 239 & 7.14 & NIST \\
5218.84 & 10.38 & -0.170 & 242 & 6.74 & NIST \\
5257.89 & 10.46 & -0.530 & 192 & 6.91 & NIST \\
5245.45 & 10.46 & -0.540 & 210 & 7.01 & NIST \\
5228.89 & 10.45 & -0.300 & 203 & 6.73 & NIST \\
5222.35 & 10.52 & -0.280 & 140 & 6.45 & NIST \\
5311.91 & 10.54 & -1.020 & 89 & 6.91 & NIST \\
5315.08 & 10.54 & -0.420 & 166 & 6.72 & NIST \\
5351.93 & 10.50 & -1.110 & 118 & 7.15 & NIST \\
5405.66 & 10.52 & -0.440 & 217 & 6.98 & NIST \\
5399.56 & 10.52 & -0.750 & 105 & 6.73 & NIST \\
5375.84 & 10.45 & -0.330 & 247 & 6.97 & NIST \\
5443.44 & 10.48 & -0.600 & 194 & 7.01 & NIST \\
5388.03 & 10.45 & -0.690 & 121 & 6.73 & NIST \\
5445.80 & 10.54 & -0.110 & 224 & 6.70 & NIST \\
5479.40 & 10.56 & -0.350 & 205 & 6.86 & NIST \\
5444.39 & 10.60 & -0.170 & 228 & 6.81 & NIST \\
5451.32 & 10.50 & -0.650 & 196 & 7.08 & NIST \\
5532.09 & 10.52 & -0.100 & 217 & 6.65 & NIST \\
5488.78 & 10.60 & -0.400 & 157 & 6.71 & NIST \\
5548.23 & 10.62 & -0.540 & 223 & 7.17 & NIST \\
5563.39 & 10.63 & -0.550 & 129 & 6.73 & NIST \\
5571.52 & 10.60 & -1.060 & 105 & 7.09 & NIST \\
4866.20 & 10.31 & -0.670 & 157 & 6.80 & NIST \\
4833.19 & 2.66 & -4.800 & 160 & 6.75 & NIST \\
4839.99 & 2.68 & -4.900 & 190 & 6.97 & NIST \\
4869.97 & 10.35 & -1.540 & 35 & 6.85 & NIST \\
4838.55 & 10.33 & -1.700 & 47 & 7.14 & NIST \\
4893.78 & 2.83 & -4.300 & 246 & 6.65 & NIST \\
4946.90 & 10.35 & -1.500 & 56 & 7.04 & NIST \\
4971.22 & 10.36 & -0.500 & 202 & 6.87 & NIST \\
4948.09 & 10.31 & -0.220 & 206 & 6.58 & NIST \\
4977.03 & 10.36 & -0.040 & 244 & 6.61 & NIST \\
4958.82 & 10.38 & -0.760 & 146 & 6.88 & NIST \\
5029.09 & 10.36 & -0.630 & 210 & 7.04 & NIST \\
4977.92 & 10.23 & -0.600 & 149 & 6.66 & NIST \\
5065.11 & 10.43 & -0.550 & 198 & 6.94 & NIST \\
5106.11 & 10.33 & -0.250 & 194 & 6.57 & NIST \\
5006.84 & 10.38 & -0.360 & 196 & 6.71 & NIST \\
5060.25 & 10.45 & -0.650 & 136 & 6.76 & NIST \\
5094.90 & 10.47 & -0.720 & 138 & 6.85 & NIST \\
5115.06 & 10.43 & -0.500 & 160 & 6.72 & NIST \\
5141.38 & 10.47 & -0.770 & 174 & 7.07 & NIST \\
5180.31 & 10.39 & -0.090 & 212 & 6.53 & NIST \\
5199.12 & 10.38 & 0.120 & 246 & 6.47 & NIST \\
5223.26 & 10.39 & -0.170 & 203 & 6.57 & NIST \\
5186.87 & 10.47 & -0.190 & 159 & 6.43 & NIST \\
5219.92 & 10.52 & -0.550 & 136 & 6.70 & NIST \\
4598.49 & 7.80 & -1.540 & 242 & 6.73 & NIST \\
4601.34 & 2.88 & -4.430 & 220 & 6.75 & NIST \\
6362.47 & 10.91 & -0.490 & 182 & 7.15 & NIST \\
6375.80 & 10.93 & -0.010 & 184 & 6.69 & NIST \\
6548.39 & 11.02 & 0.240 & 197 & 6.58 & NIST \\
6541.37 & 11.05 & 0.350 & 192 & 6.46 & NIST \\
6650.98 & 7.13 & -2.400 & 115 & 6.68 & NIST \\
6357.16 & 10.91 & 0.240 & 217 & 6.59 & NIST \\
7616.93 & 11.27 & -0.820 & 98 & 7.32 & NIST \\
  &   &   &   &   &   \\
Mean: &   &   &   & 6.78 $\pm$ 0.19 &   \\
  &   &   &   &   &   \\
Co II &   &   &   &   &   \\
4516.65 & 3.46 & -2.460 & 156 & 4.66 & NIST \\
4569.26 & 3.41 & -2.400 & 183 & 4.68 & NIST \\
4660.65 & 3.36 & -2.350 & 242 & 4.81 & NIST \\
  &   &   &   &   &   \\
Mean: &   &   &   & 4.72 $\pm$ 0.08 &   \\
  &   &   &   &   &   \\
Ni I &   &   &   &   &   \\
4904.41 & 3.54 & -0.170 & 25 & 5.47 & NIST \\
5081.11 & 3.85 & 0.300 & 50 & 5.49 & NIST \\
4714.42 & 3.38 & 0.230 & 66 & 5.45 & NIST \\
  &   &   &   &   &   \\
Mean: &   &   &   & 5.47 $\pm$ 0.02 &   \\
  &   &   &   &   &   \\
Ni II &   &   &   &   &   \\
4192.07 & 4.01 & -3.060 & 254 & 5.58 & K88 \\
5003.41 & 12.54 & 0.702 & 105 & 5.55 & K88 \\
  &   &   &   &   &   \\
Mean: &   &   &   & 5.56 $\pm$ 0.03 &   \\
  &   &   &   &   &   \\
Ge II &   &   &   &   &   \\
7049.37 & 8.08 & 0.000 & Synth & 3.60 & NIST \\
7145.39 & 8.06 & -0.300 & Synth & 3.77 & NIST \\
6021.04 & 7.74 & -0.040 & Synth & 3.80 & NIST \\
  &   &   &   &   &   \\
Mean: &   &   &   & 3.72 $\pm$ 0.09 &   \\
  &   &   &   &   &   \\
Sr II &   &   &   &   &   \\
4215.52 & 0.00 & -0.166 & 716 & 3.39 & NIST \\
4077.71 & 0.00 & 0.148 & 707 & 3.12 & NIST \\
4305.45 & 3.04 & -0.100 & 360 & 3.55 & NIST \\
4161.80 & 2.94 & -0.470 & 270 & 3.56 & NIST \\
  &   &   &   &   &   \\
Mean: &   &   &   & 3.40 $\pm$ 0.20 &   \\
  &   &   &   &   &   \\
Y II &   &   &   &   &   \\
5728.89 & 1.84 & -1.120 & 62 & 2.73 & NIST \\
5402.77 & 1.84 & -0.510 & 220 & 2.84 & NIST \\
4786.58 & 1.03 & -1.290 & 110 & 2.74 & NIST \\
4854.86 & 0.99 & -0.380 & 314 & 2.49 & NIST \\
4900.12 & 1.03 & -0.090 & 370 & 2.38 & NIST \\
5087.42 & 1.08 & -0.170 & 402 & 2.56 & NIST \\
5205.72 & 1.03 & -0.350 & 318 & 2.45 & NIST \\
4309.62 & 0.18 & -0.747 & 257 & 2.24 & NIST \\
4398.01 & 0.13 & -0.999 & 280 & 2.51 & NIST \\
4682.32 & 0.41 & -1.509 & 130 & 2.66 & NIST \\
3950.35 & 0.10 & -0.488 & 370 & 2.39 & NIST \\
5662.92 & 1.94 & 0.160 & 324 & 2.51 & NIST \\
  &   &   &   &   &   \\
Mean: &   &   &   & 2.54 $\pm$ 0.17 &   \\
  &   &   &   &   &   \\
Zr II &   &   &   &   &   \\
4208.99 & 0.71 & -0.510 & 386 & 2.79 & Ljung \\
4359.74 & 1.24 & -0.510 & 261 & 2.68 & Ljung \\
4440.45 & 1.21 & -1.040 & 238 & 3.11 & Ljung \\
4495.44 & 1.21 & -1.830 & 75 & 3.21 & Ljung \\
4379.78 & 1.53 & -0.356 & 334 & 2.95 & CC \\
4308.94 & 1.49 & -0.800 & 142 & 2.73 & Kurucz \\
4494.41 & 2.41 & -0.230 & 290 & 3.22 & Malcheva \\
4496.96 & 0.71 & -0.890 & 270 & 2.72 & Ljung \\
4613.95 & 0.97 & -1.540 & 156 & 3.15 & Ljung \\
4661.78 & 2.41 & -0.800 & 115 & 3.15 & Kurucz \\
3998.98 & 0.56 & -0.520 & 410 & 2.86 & Ljung \\
4050.32 & 0.71 & -1.060 & 250 & 2.91 & Ljung \\
4156.24 & 0.71 & -0.780 & 270 & 2.67 & Ljung \\
4211.88 & 0.53 & -1.040 & 340 & 3.03 & Ljung \\
4018.38 & 0.96 & -1.270 & 137 & 2.88 & Ljung \\
5350.10 & 1.83 & -0.390 & 261 & 2.81 & Ljung \\
  &   &   &   &   &   \\
Mean: &   &   &   & 2.93 $\pm$ 0.19 &   \\
  &   &   &   &   &   \\
Ba II &   &   &   &   &   \\
6141.71 & 0.70 & -0.032 & 350 & 2.35 & NIST \\
6496.90 & 0.60 & -0.407 & 280 & 2.45 & NIST \\
4934.08 & 0.00 & -0.160 & 302 & 2.00 & NIST \\
4554.03 & 0.00 & 0.140 & 409 & 2.12 & NIST \\
  &   &   &   &   &   \\
Mean: &   &   &   & 2.23 $\pm$ 0.21&   \\
\enddata
\tablenotetext{a}{Normalized such that log $\sum\mu_i\epsilon(i)$ = 12.15.}
\tablenotetext{b}{Sources of log \textit{gf} values.}
\tablerefs{NIST: \citet{NIST}; K88: \citet{K88}; CC: \citet{CC}; FMW: \citet{FMW}; GARZ: \citet{GARZ}; Luck: Compilations by R. E. Luck; Tom97: \citet{Tom97}; KP: \citet{KP}; NBS: \citet{nbs}; Kurucz: Kurucz Database; Malcheva: Malcheva et al. (2006); Ljung: Ljung et al. (2006)}
\end{deluxetable}
\section{Spectrum synthesis of Hydrogen line}\label{hsyntheis}

\begin{figure}[h!]
\includegraphics[width=\hsize]{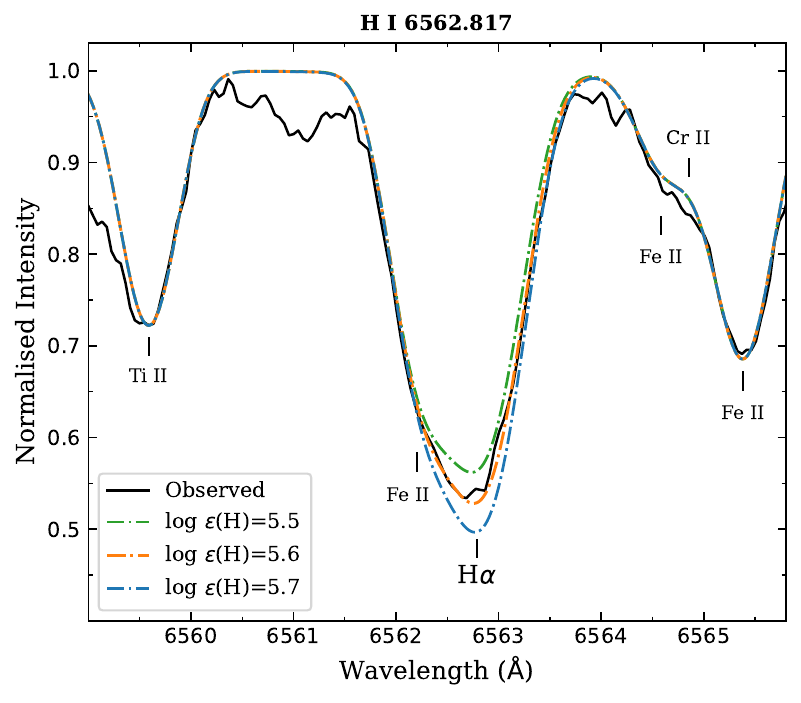}
\caption{Observed and synthesized H$\alpha$ line profile is shown for A\,980.
\label{fig:Hsyntheis}}
\end{figure}

The available hydrogen Balmer lines (H$\alpha$, H$\beta$, and H$\gamma$) are synthesized to obtain the hydrogen abundance (log $\epsilon$(H)). For example, synthesis of the H$\alpha$ line is shown in Figure \ref{fig:Hsyntheis}. The hydrogen Balmer lines are synthesized using the optimum model atmosphere with T$_{\text{eff}}$ = 9250 K, log g = 1.25 cgs and $\xi$ = 11.5 km s$^{-1}$, and C/He=0.3 \%.
\clearpage
\bibliography{A980EHe}{}
\bibliographystyle{aasjournal}

\end{document}